# Hysteresis in the Conductance of Asymmetrically Biased GaAs Quantum Point Contacts with in-plane Side Gates

N. Bhandari[1], M. Dutta[1], J. Charles[1], M. Cahay[1,2], and R. S. Newrock[2]

[1] School of Electronics and Computing Systems
University of Cincinnati, Cincinnati, Ohio 45221, USA
[2] Physics Department, University of Cincinnati, Cincinnati, Ohio 45221, USA

**Abstract**

We have observed hysteresis between the forward and reverse sweeps of a common mode bias applied to the two in-plane side gates of an asymmetrically biased GaAs quantum point contact (QPC). The size of the hysteresis loop increases with the amount of bias asymmetry $\Delta V_g$ between the two side gates and depends on the polarity of $\Delta V_g$. Our results are in qualitative agreement with Non-Equilibrium Green's Function simulations including the effects of dangling bond scattering on the sidewalls of the QPC. It is argued that hysteresis may constitute another indirect proof of spontaneous spin polarization in the narrow portion of the QPC.

1. Introduction

Experimental observations of hysteresis in the transport properties of two-dimensional electron gas (2DEG) made of single layer and double layer quantum well heterostructures used in the study of the integer and fractional quantum Hall effects have been widely reported [1-7]. In the double-layer structure, the origin of the hysteresis was attributed to a phase transition between oppositely polarized ground states localized in the different layers [1]. In the study of the fractional quantum Hall effect, the hysteretic behavior in the modulation-doped heterostructures was linked to the onset of two-dimensional ferromagnetism resulting from the competition of spin polarized and a spin unpolarized ground states [2,3,7]. Hysteresis between the forward and reverse sweeps of an external magnetic field in the transport properties of a quantum wire operating in the integer quantum Hall regime was predicted by Ihnatsenka and Zozoulenko [8]. This hysteresis was shown to be due to the coexistence of two different ground states associated with spatially spin polarized and spatially spin unpolarized edge channels in the vicinity of the wire boundaries over a specific range of the applied magnetic field. Ihnatsenka and Zozoulenko also showed that the hysteretic behavior is absent for steep and smooth confining potentials in the quantum wire and is present only for a limited range of intermediate confinement slopes. In fact, the hysteresis behavior in the integer quantum Hall effect cannot be explained using a single electron model but has a many-body origin resulting from the non-linear intricate interplay between the confining potential, the Coulomb interaction and the exchange interaction in the quantum wire [8].

Following the pioneering work of Thomas et al. [9], there have been many experimental reports of anomalies in the quantized conductance of QPCs appearing at non-integer multiples of $G_0$ $(=2e^2/h)$, including $0.25G_0$, $0.5G_0$, and $0.7G_0$ [10-12]. Thomas et al. were the first to suggest that the 0.7$G_0$ conductance anomaly was an indirect evidence of a spontaneous ferromagnetic spin polarization in the narrow portion of the QPC. Many theories of conductance anomalies have now been developed on that assumption, including a spontaneous spin polarization in the narrow portion of the QPC due to the exchange-correlation interaction [13-15], the formation of quasi-bound states [16], and the Kondo effect [17], among others [10-12]. There is now an increasing number of reports showing that the number and location of conductance anomalies can be controlled by creating an asymmetry in the QPC's electrostatic confining potential [18-21].

The existence of a spontaneous ferromagnetic spin polarization in QPCs should be accompanied by the presence of hysteresis loop(s) in the conductance curves for appropriate device parameters and biasing conditions. There is however only one experimental report by Shailos et al. showing hysteresis in the conductance plots of QPCs [18]. They investigated the linear transport properties of QPCs while deliberately breaking the symmetry of its confining potential in a controlled manner [18]. Their devices consisted of a conventional split-gate QPC modified with an additional perturbing finger gate used to modulate the electron density on one side of the device. As the bias applied to this finger gate was varied, Shailos et al. observed several reproducible conductance anomalies below the last integer plateau $G_0$ together with substantial modifications of the integer-plateau staircase. The conductance curves showed a pronounced hysteresis as the bias on the split-gate was swept in opposite directions. Shailos et al. suggested that several extrinsic mechanisms that could cause the hysteresis, including surface states at the metal–semiconductor junction and/or impurity states within the heterostructure. They reached the important conclusion that hysteresis in the conductance curves of QPCs could be used as a tool to evaluate the sensitivity of the QPC transport to microscopic configuration changes.

2. Experimental Results

In this work, we report the observation of hysteresis in the conductance of GaAs QPCs while sweeping up and down a common gate signal applied to its two asymmetrically biased in-plane side gates (SGs). These results complement our earlier report the observation of a robust anomalous conductance plateau near $G = 0.5\ G_0$ in asymmetrically biased AlGaAs/GaAs QPCs with in-plane SGs in the presence of lateral spin-orbit coupling which was interpreted as evidence of spin polarization in the narrow portion of the QPC [22]. Hereafter, we show that the size of the hysteresis loop increases with the amount of bias asymmetry $\Delta V_g$ between the two SGs and depends on the polarity of $\Delta V_g$. Our results are in qualitative agreement with Non-Equilibrium Green's Function (NEGF) simulations [23,24,25] modified to include the effects of dangling bond scattering on the sidewalls of the QPC [26].

We used a 2DEG formed at the hetero-interface of Si modulation doped GaAs/AlGaAs quantum heterostructure to fabricate the QPC device. The details of the heterostructure layers are are the same as in ref.[27]. The 2DEG was characterized by Shubnikov-de Haas (SdH) and quantum Hall measurements; its carrier density and mobility were found to be $1.6 \times 10^{11}/cm^2$ and $1.9 \times 10^5\ cm^2/V\text{-}s$, respectively. Sample cleaning and device fabrication procedure has been described elsewhere [27, 28].

In the device reported here, the narrow portion of the QPC channel has a width (along y-direction) and length (along x-direction) around 350 nm and 400 nm, respectively (Fig.1). The electrostatic width of the QPC channel was changed by applying bias voltages to the metallic in-plane SGs, depleting the channel near the side walls of the QPC. Battery operated DC voltage sources were used to apply constant voltages $V_{G1}$ and $V_{G2}$ to the two gates. An asymmetric potential $\Delta V = V_{G1}-V_{G2}$ between the two gates was applied to create spin polarization in the channel. The QPC conductance was then recorded as a function of a 3 mHz *cyclical* sweep voltage, Vsweep, applied to the two gates in addition to the potentials $V_{G1}$ and $V_{G2}$. The linear conductance G (=I/V) of the channel was measured for different $\Delta V_G$ as a function of Vsweep, using a four-probe lock-in technique with a drive frequency of 17 Hz and a drain-source drive voltage of 100 μV. Leakage measurements were done beforehand to ensure that the gates are not leaking. Because GaAs has a large surface depletion as a result of Fermi level pining by surface states [28], a large positive potential (about 12 V) was needed on both gates to obtain a conducting channel at T=4.2K. The potential on both gates was then gradually reduced in the range of a few volts making sure the channel remained open.

Figure 2 shows the conductance of the QPC as a function of the sweep voltage $V_{sweep}$ for different negative asymmetric biases ($\Delta V_G=V_{G1}-V_{G2}$) between the gates. The potential applied to gate G1 is fixed at 0 V. The potential on gate G2 is, from left to right, is set equal to 0, 0.6, 1.2, 1.8, 2.4, 3.0, 3.6, and 4.2 V, respectively. In Fig.2, the second to last conductance curves have been shifted to the right for clarity. The conductance curves shows two anomalous conductance plateaus, one slightly about $0.5G_0$ and another around $1.25G_0$, whose exact locations depend on the value of $\Delta V_G$. Furthermore, all conductance curves show hysteresis between the forward and reverser sweeps of the common mode signal $V_{sweep}$. The hysteresis can be seen even for the case of $\Delta V_G = 0$ and increases with $\Delta V_G$.

Figure 3 shows the conductance of the QPC as a function of the sweep voltage $V_{sweep}$ for both positive and negative values of the asymmetric biases ($\Delta V_G=V_{G1}-V_{G2}$) between the SGs. The potential applied to gate G1 is fixed at 1.5 V. The potential on gate G2 is, from left to right, 0.3, 0.9, 1.5, 2.1 and 2.7 V. This corresponds to a range of $\Delta V_G$ from -1.2 to 1.2 V. Figure 3 shows conductance anomalies similar to those shown in Fig.2. In addition, the hysteresis loop is the smallest for $\Delta V_G = 0$ V and increases as $\Delta V_G$ is made either more positive or negative. For a given absolute value of $|\Delta V_G|$, the size of the hysteresis loop is different when $V_{G1}-V_{G2} = |\Delta V_G|$ or $V_{G1}-V_{G2} = -|\Delta V_G|$.

3. Discussion

Recently, we reported a theoretical investigation of the onset of hysteresis of asymmetrically biased GaAs QPC with in-plane side gates in the presence of lateral spin-orbit coupling (LSOC) [23]. We showed that the hysteresis in the conductance versus common gate bias applied to the two side gates exists only if the narrow portion of the QPC is long enough. The hysteresis is absent if the effects of electron-electron interaction are neglected and increases with the strength of the electron-electron interaction. The hysteresis appears in the region of conductance anomalies, i.e., less than $2e^2/h$, and is due to multistable spin textures in these regions [23]. Hereafter, we used the NEGF approach to investigate the influence of dangling bond scattering on the conductance of a GaAs with in-plane side gates which has been shown to have a strong influence on the location of the conductance anomalies [26].

The model QPC we have used is shown in Fig. 4, where the white region represents the QPC channel with openings at the ends. The gray area represents the etched isolation trenches that define the lithographic dimensions of the QPC constriction. The black strips show the four contact electrodes connected to the QPC device: source, drain and two SGs. Symmetric and asymmetric SG voltages can be applied. We consider the QPC in Fig. 4 to be made from a nominally symmetric GaAs quantum well (QW). Both the effects of Rashba and Dresselhaus spin-orbit interaction were neglected. The only spin-orbit interaction considered is the LSOC due to the lateral confinement of the QPC channel, provided by the isolation trenches and the bias voltages of the side gates [29]. The free-electron Hamiltonian of the QPC is given by

$$H = H_0 + H_{SO}$$
$$H_0 = \frac{\hbar^2(k_x^2 + k_y^2)}{2m^*} + U(x,y) \quad (1)$$
$$H_{SO} = \beta\vec{\sigma}\cdot(\vec{k}_x \times \vec{\nabla} U(x,y))$$

In equation (1), $H_{SO}$ is the LSOC interaction term, $\beta$ the intrinsic SOC parameter, $\vec{\sigma}$ the vector of Pauli spin matrices, and $\vec{B}_{SO} = \beta(\vec{k}_x \times \vec{\nabla} U(x,y))$ is the effective magnetic field induced by LSOC. The effective mass in the GaAs channel was set equal to $m^* = 0.07 m_0$, where $m_0$ is the free electron mass. The 2DEG is assumed to be located in the $(x, y)$ plane, $x$ being the direction of current flow from source to drain and $y$ the direction of transverse confinement of the channel. $U(x, y)$ is the confinement potential, which includes the potential introduced by gate voltages and the conduction band discontinuity at the GaAs/air interface.

The conductance of the QPC was calculated using a NEGF method under the assumption of ballistic transport [24, 25]. We used a Hartree-Fock approximation following Lassl et al. [30] to include the effects of electron-electron interaction in the QPC. More specifically, the electron-eleectron interaction was taken into account by considering a repulsive Coulomb contact potential, $V_{int}(x,y;x',y') = \gamma\, \delta(x-x')\, \delta(y-y')$, where $\gamma$ indicates the electron-electron interaction strength. As a result, an interaction self-energy, $\Sigma_{int}^{\sigma}(x, y)$, must be added to the Hamiltonian in Eq.(1). At the interface between the rectangular region of size $w_2 \times l_2$ and vacuum, the conduction band discontinuities at the bottom and the top interface were modeled using a smooth conductance band change spread over a range $d$ selected to be in the nm range to represent a gradual variation of the conduction band profile from the inside of the quantum wire to the vacuum region [23-25]. A similar grading was also used along the walls going from the wider part of the channel to the central constriction of the QPC (Fig.4). This gradual change in the conduction band at the QPC channel/vacuum interface is responsible for the LSOC that triggers the spin polarization of the QPC in the presence of an asymmetry in $V_{sg1}$ and $V_{sg2}$ [24]. The parameter $d$ used in the potential profile modeling the conduction band variation at the channel/vacuum interface was set equal to 1.6 nm. The conductance of the QPC was then calculated using the NEGF with a non-uniform grid configuration containing more grid points at the interface of the QPC with vacuum. All calculations were performed at a temperature $T = 4.2$ K. For a dangling bond located at location $(x_1, y_1)$, we model its potential energy in the 2 DEG as follows,

$$U_{impurity}(x,y) = \frac{q^2}{4\pi\varepsilon_0\varepsilon_r\sqrt{(x-x_1)^2+(y-y_1)^2+\Delta^2}}, \quad (2)$$

where $\Delta = \frac{q}{4\pi\varepsilon_0\varepsilon_r U_0}$, and $U_0$ is the maximum strength of the impurity potential.

We used $\varepsilon_r = 12.9$, the relative dielectric constant of GaAs, and $U_0$ was set equal to 200 meV. The dangling bond has coordinates $(x_1, y_1) = (\frac{l_1-l_2}{2} + \frac{l_2}{4}, \frac{w_1-w_2}{2} + \frac{d}{2})$, i.e., it is located $1/4^{th}$ of the way from the left side of the narrow portion of the QPC and in the middle of the bottom side wall interface. Figure 4 shows a plot of the conductance G (in units of $e^2/h$) of the GaAs QPC as a function of the common mode signal $V_{sweep}$ applied to the two SGs. The solid and dashed curves correspond to the forward and reverse sweeps, respectively. The two different set of curves labeled I and II correspond to the biasing conditions on the gates: (I) $V_{sg1} = 0.2V + V_{sweep}$ and $V_{sg2} = -0.2V + V_{sweep}$ and (II) $V_{sg1} = -0.2V + V_{sweep}$ and $V_{sg2} = 0.2V + V_{sweep}$. The temperature is set equal to 4.2K and the device dimensions are $l_2 = 32$nm, $l_1 = l_2 + 32$nm, $w_2 =$ 16nm, and $w_1$ = 48nm. The following parameters were used: $V_{ds}$ = 0.1mV, T = 4.2K, $\gamma$ = 3.7 in units of $\hbar^2/2m^*$, and $\beta$ = 5 Å$^2$. Also shown in Fig.4 are the conductance curves calculated with no dangling bond present (curves labeled *"No Impurity"*). In this case, the conductance curves are the same for the two biasing configurations I and II.

Figure 4 shows that hysteresis appears in the range $V_{sweep}$ where conductance anomalies occur. In qualitative agreement with the experimental results of Figures 2 and 3, a hysteresis is present even for the case where $\Delta V_G$ = 0. In fact, Hsiao et al. have shown that a finite spin polarization (an associated hysteresis according to our simulations) can be created in the vicinity of a sufficiently long symmetrically biased QPC with top gates in the presence of spin-orbit interaction [31,32]. In Fig.4, the hysteresis loops appear over a larger range of $V_{sweep}$ for $\Delta V_g \neq 0$ and are quite different for opposite polarities of the bias asymmetry between the SGs, in qualitative agreement with our experimental results. Even though not shown here, the NEGF simulations predict identical hysteresis loops for opposite polarities when $\Delta V_g \neq 0$ with no dangling bond present (in this case, the contributions from the up- and down-spin electrons to the conductance are interchanged when the polarity is flipped but the total conductance for both the forward and reverse sweeps stay the same). The difference of the hysteresis loops for opposite polarities of a finite $\Delta V_g$ is therefore another indirect evidence of the importance of dangling bond scattering in QPCs with in-plane SGs [26]. Even though not shown here, we have found that the experimental hysteresis loops are also dependent on the sweep rates of the common mode signal used, another indirect evidence of the different time scales involved in the charging and discharging of dangling bonds and impurities in close proximity to the narrow portion of the QPC.

The experimental hysteresis is observed over the entire range of common mode signal applied to the SGs whereas several hysteresis loops are predicted in the numerical simulations. A better quantitative agreement could be reached by performing simulations for structures with dimensions closer to the experimental ones and also by including the effects of other scattering mechanisms, such as surface roughness scattering [26] and scattering from remote impurities used for modulation doping of the 2DEG.

**Acknowledgment**

This work is supported by NSF Award ECCS 1028423. James Charles acknowledges support under NSF-REU award 007081.

**Figures**

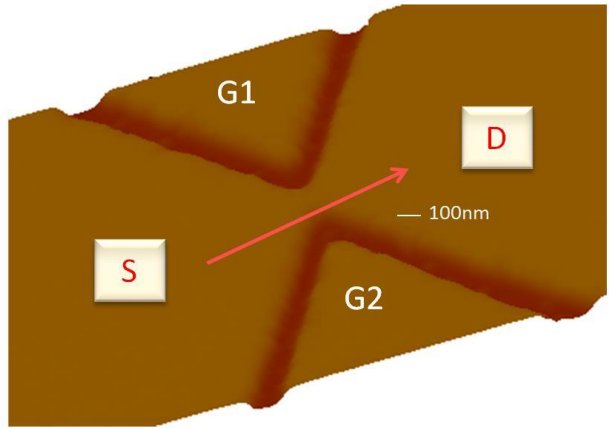

**Fig. 1:** Three-dimensional AFM image of a QPC with two side gates (G1 and G2) fabricated with chemical wet-etching technique. An asymmetric Lateral Spin Orbit Coupling (LSOC) is generated using an asymmetric bias between the two side gates. The narrow portion of the QPC has a width and length around 350 nm and 275 nm, respectively.

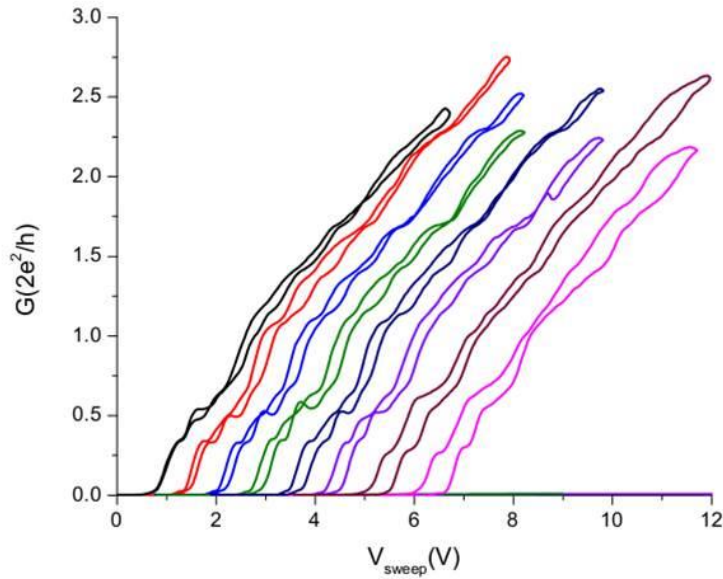

**Fig. 2:** Conductance of a QPC (in units of $2e^2/h$) as a function of the sweep voltage $V_{sweep}$ applied to the side gates. The measurements include forward and reverse sweeps of $V_{sweep}$.

The sweep voltage is superimposed on the potentials $V_{G1}$ and $V_{G2}$ applied to the gates to create an asymmetry. The potential applied to gate G1 is fixed at 0 V. The potential on gate G2 is, from left to right, 0, 0.6, 1.2, 1.8, 2.4, 3.0, 3.6, and 4.2 V.

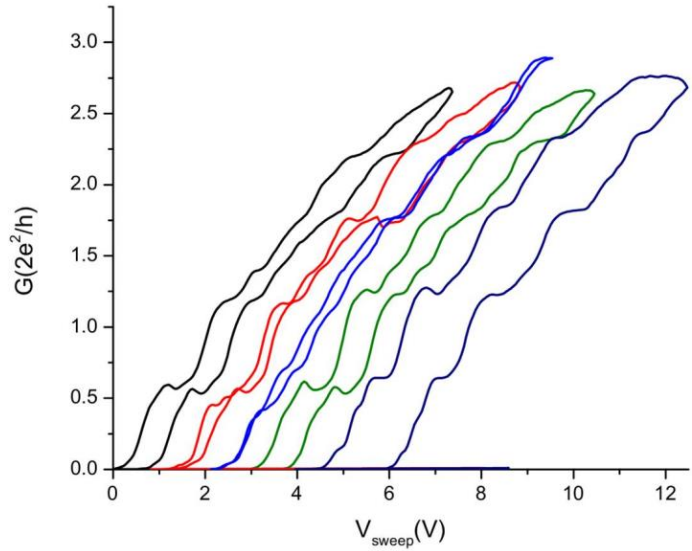

**Fig. 3:** Conductance of a QPC (in units of $2e^2/h$) as a function of the sweep voltage $V_{sweep}$ applied to the side gates. The measurements include forward and reverse sweeps of $V_{sweep}$. The sweep voltage is superimposed on the potentials $V_{G1}$ and $V_{G2}$ applied to the gates to create an asymmetry. The potential applied to gate G1 is fixed at 1.5 V. The potential on gate G2 is, from left to right, 0.3, 0.9, 1.5, 2.1 and 2.7 V.

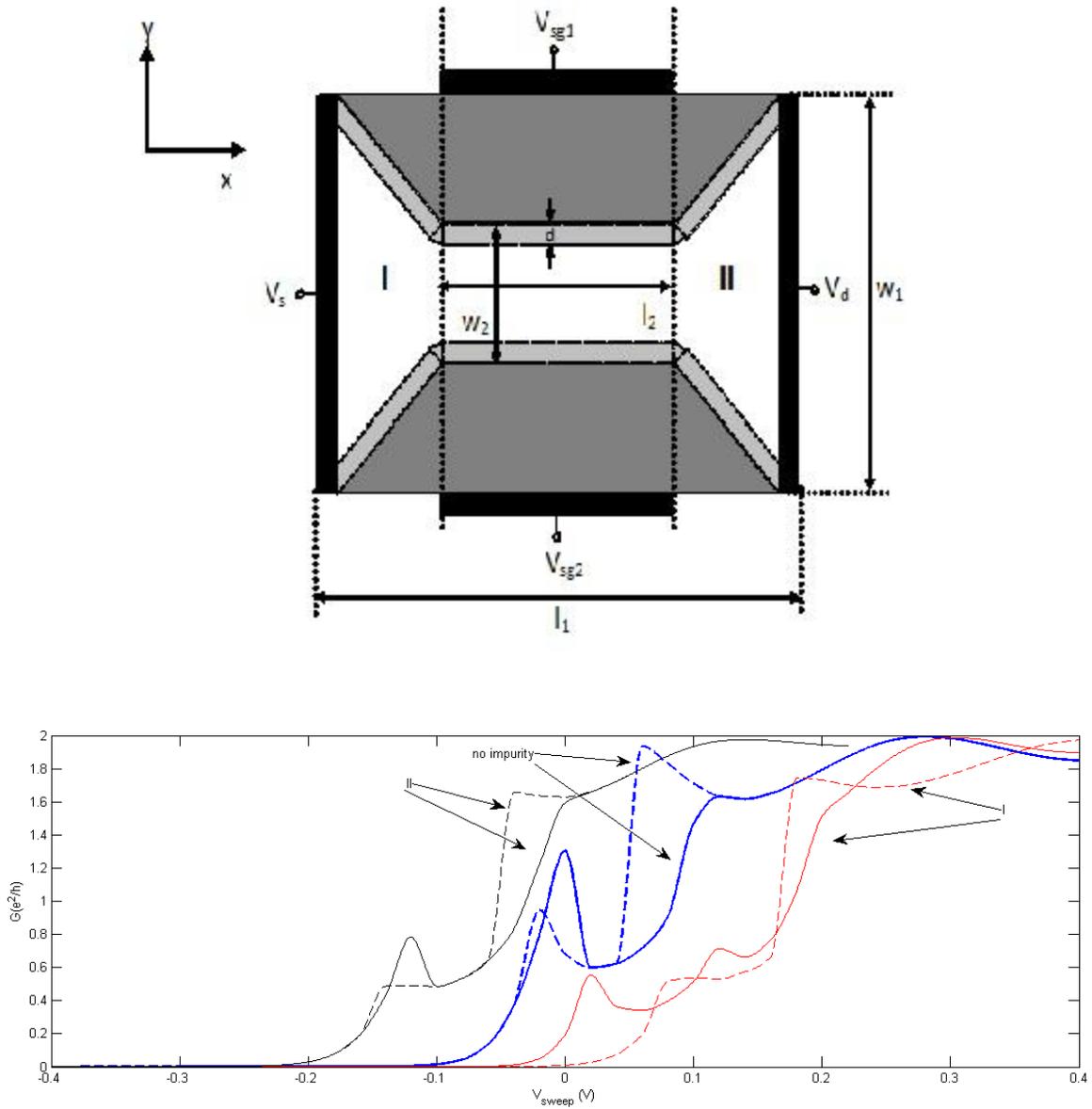

**Fig. 4: (Top)** Schematic illustration of the QPC geometrical layout used in the NEGF simulations. **(Bottom)** Influence of dangling bond scattering on the hysteresis of the conductance $G$ (in units of $e^2/h$) of a GaAs QPC calculated as a function of the common mode signal $V_{sweep}$ applied to the two in-plane SGs. The solid and dashed curves correspond to the forward and reverse sweeps, respectively. The two different set of curves labeled I and II correspond to the biasing conditions

on the gates: (I) $V_{sg1} = 0.2V + V_{sweep}$ and $V_{sg2} = -0.2V + V_{sweep}$ and (II) $V_{sg1} = -0.2V + V_{sweep}$ and $V_{sg2} = 0.2V + V_{sweep}$. The temperature is set equal to 4.2K and the device dimensions are $l_2 = 32$nm, $l_1 = l_2 + 32$nm, $w_2 = 16$nm, and $w_1 = 48$nm. The following parameters were used: $V_{ds} = 0.1$mV, T = 4.2K, $\gamma = 3.7$ in units of $\hbar^2/2m^*$, and $\beta = 5$ Å$^2$. A dangling bond is located in the narrow portion of the QPC at $(x_1, y_1) = (\frac{l1-l2}{2} + \frac{l2}{4}, \frac{w1-w2}{2} + \frac{d}{2})$ with strength equal to 200 meV. Also shown for comparisons are the conductance curves calculated without a dangling bond present (curves labeled "*No impurity*"). The curves for the biasing configuration II have been shifted by -0.1V for clarity.